\documentclass[a4paper,12pt]{article}

\usepackage[T1]{fontenc}
\usepackage[english]{babel}
\usepackage[margin=2.0cm]{geometry}
\usepackage{amsmath, amssymb,booktabs}

\newcommand{\At}{\A_{\tau}}
\newcommand{\gor}{\;\;\big|\;\;}
\newcommand{\Nilp}{{0}}
\newcommand{\A}{\mbox{$\mathbb{A}$}}
\newcommand{\nat}{\mathbb{N}}

\newcommand{\nar}[2]{\xrightarrow{#1}_{#2}}

\newcommand{\fase}{\texttt{FASE}}

\newcommand{\fifo}{\mbox{{\sf Fifo}}}
\newcommand{\pipe}{\mbox{{\sf Pipe}}}
\newcommand{\buff}{\mbox{{\sf Buff}}}
\newcommand{\mem}{\mbox{{\sf Mem}}}

\newcommand{\bc}{\mbox{{\sf BC}}}
\newcommand{\Pp}{\mbox{$\mathbb{P}$}}

\newcommand{\RT}{\mbox{\sf RT}}
\newcommand{\DL}{\mbox{\sf DL}}
\newcommand{\RTS}{\mbox{\sf RTS}}
\newcommand{\RRTS}{\mbox{\sf rRTS}}
\newcommand{\Chi}{\text{\large\raisebox{0.45ex}{$\chi$}}}
\newcommand{\institute}[1]{\\{\scriptsize
   \begin{tabular}[t]{@{\small}c@{}}#1\end{tabular}}}
\newcommand{\email}[1]{\\{\footnotesize\tt #1}}

\title{A Framework for the Evaluation of Worst-Case System Efficiency
	\author{M. Callisto De Donato, M.R. Di Berardini
		\institute{Scuola di Scienze e Tecnologie, Sezione Informatica. Universit\`a di Camerino}
		\email{\{massimo.callisto, mariarita.diberardini\}@unicam.it}
		 }
	\date{}
}
\begin{document}
\maketitle

\begin{abstract}
In this paper we present \fase\ (\textbf{F}ast \textbf{A}synchronous \textbf{S}ystems
\textbf{E}valuation), a tool for evaluating worst-case efficiency of asynchronous systems. This tool
implements some well-established results in the setting of a timed CCS-like process algebra: PAFAS
(a Process Algebra for Faster Asynchronous Systems).  Moreover, we discuss some new solutions that
are useful to improve the applicability of \fase\ to concrete meaningful examples. We finally use
\fase\ to evaluate the efficiency of three different implementations of a bounded buffer and compare
our results with previous ones obtained when the same implementations have been contrasted according
to an efficiency preorder.
\end{abstract}

\section{Introduction}
In concurrent and distributed systems, study time aspects at an early stage of software
development plays an important role to ensure the correct temporal execution of system activities. 
In recent years, PAFAS has been proposed as a powerful tool for evaluating
the worst-case efficiency of asynchronous systems \cite{CVJ00,CV05}. PAFAS is a CCS\cite{Mil89}-like timed
process algebra where system activities are represented by durationless actions and time passes in
between them \cite{ABC09}. Thus, actions are atomic and instantaneous but have associated a time
bound interpreted as the maximal time delay for their execution. This timing information can be used
to evaluate efficiency without influence functionality (which actions are performed). So, compared to CCS, also PAFAS
treats the full functionality of asynchronous systems.
In~\cite{CVJ00}, process are compared via a variant of the testing approach developed in~\cite{DH84} by 
considering test environments (as in~\cite{DH84}) together with a time bound. A process is embedded
into the environment (via parallel composition) and satisfies a  test if success is reached before
the time bound in {\em every} run of the composed system, i.e.\ {\it even in the worst case}. This
gives rise to a faster-than preorder relation over processes that is naturally an {\it efficiency
preorder}. Furthermore, this preorder can be characterised as inclusion of a special kind of refusal
traces which provide decidability of the testing preorder for finite state processes.
The faster-than preorder has been equivalently defined in~\cite{CV05} on the basis of a performance
function that gives the worst-case time needed to satisfy any test environment (or user behaviour).
Another key result in~\cite{CV05} shows that, whenever the above testing scenario is adapted by
considering only test environments that want $n$ task to be performed as fast as possible, the
performance function is {\em asymptotically linear}. This function is a {\em quantitative}
performance measure that describes how fast a  system responds to requests from the environment.
This paper presents \fase, a corresponding tool that supports us to automatically evaluate the
worst-case performance of a PAFAS process. \fase\ has been successfully used in~\cite{BCC+09} to
relate three different implementations of bounded buffer: \fifo\
(first-in-first-out queue), \pipe\ (sequence of cells connected end-to-end) and \buff\ (an array
used in a circular fashion). The results obtained in~\cite{BCC+09} were also compared with those 
in~\cite{CDV01} where the same implementations have been contrasted via the efficiency preorder
in~\cite{CVJ00}.

\section{PAFAS}
We adopt the following notation: $\A$ (ranged over by $a,b,c,\dots$) is an infinite set of basic
actions with the special action $\omega$ reserved for observes (test processes) in the testing
scenario to signal the success of a test. Action $\tau$ represents an internal activity unobservable
for other components; we define $\At=\A\cup\{\tau\}$ where elements are ranged over by
$\alpha,\beta,\cdots$. We assume that actions in $\At$ can let time $1$ pass as maximal delay before
their execution; after that time they become {\em urgent}. The set of urgent actions is denoted by
$\underline{\A}_\tau =\{\underline{a} \,|\, a \in \A\} \cup \{\underline{\tau}\}$ and is ranged over
by $\underline{\alpha},\underline{\beta},\dots$. $\Chi$ (ranged over by $x,y,z,\dots$) is the set of
process variables, used for recursive definitions. A {\it general relabelling function} 
$\Phi:\At \to \At$ is such that $\{\alpha\in\At \,|\, \emptyset \neq
\Phi^{-1}(\alpha)\ne\{\alpha\}\}$ is finite and $\Phi(\tau)=\tau$. General relabelling functions 
subsume both relabelling and hiding (see~\cite{CVJ00}).

The set $\Pp$ of {\it (timed) processes} is the set of closed (i.e., without free variables) and
guarded (i.e., variable $x$ in a recursive term $\mu x. P$ only appears within the scope of a
action-prefix) terms generated by the following grammar: $P  ::= \Nilp \gor \gamma . P \gor P+P \gor
P\|_A P \gor P[\Phi] \gor x \gor \mu x.P$, where $\gamma$ is either $\alpha$ or $\underline{\alpha}$
for some $\alpha \in \At$, $\Phi$ is a general relabelling function, $x\in\Chi$ as expected and
$A\subseteq\A$ possibly infinite. $\Nilp$ is the Nil-process which cannot perform any action but may
let time pass without limit. $\alpha.P$ and $\underline{\alpha}.P$ is the (action-) prefixing, known
from CCS; $a .P$ can either perform $a$ immediately or can idle for time 1 and become
$\underline{a}.P$. In the latter case, the idle-time has elapsed and $a$ must either occur or be
deactivated (in a choice-context) before time may pass further. Our processes are {\em patient}: as
a stand-alone process, $\underline{a} .P$ has no reason to wait;  but as a component in
$\underline{a} .P\|_{\{a\}} a.Q$, it has to wait for synchronisation on $a$ and this can take up to
time $1$, since component $a.Q$ may idle this long. $P_1 + P_2$ models the choice between two
processes $P_1$ and $P_2$.  $P_1\|_A P_2$ is the parallel composition of two processes $P_1$ and
$P_2$ that run in parallel and have to synchronise on $A$ \cite{Hoa85}.

The temporal behaviour is given by means of the so-called \textit{refusal traces}. Intuitively, a
refusal trace records, along a computation, which actions $P$ can perform ($P \nar{\alpha}{r} P'$,
$\alpha \in \At$) and which actions $P$ can refuse to perform
($P\nar{X}{r} P'$, $X \subseteq \A$).\footnote{We omit here the (almost standard) SOS-rules defining
the transition relations $\nar{\alpha}{}$ and $\nar{X}{r}$ (see~\cite{CVJ00} for further details).}
A transition $P\nar{X}{r} P'$ is  a {\it conditional time step}. Actions in $X$ are not urgent and,
hence, $P$ is justified in not performing them and performing a time step instead. Since other
actions might be urgent, $P$ might actually be unable to refuse any possible action (e.g.\
$\underline{a}.P$ can never refuse $a$). Nevertheless, as a components of a larger
system, it can refuse some of its urgent actions due to synchronisation with the environment. As
an example: as a component of $\underline{a}.P\|_{\{a\}} a.Q$, $\underline{a}.P$ can refuse $a$
since its synchronisation partner $Q$ can do so.
 We say that $P$ perform a {\em full time step} (written $P\nar{1}{} P'$) if $P\nar{\A}{r} P'$. A
{\em discrete trace} is any sequence in $v \in (\At \cup \{1\})^*$ that $P$ can perform.
Finally, $\DL(P)$ and $\RT(P)$ are the sets of discrete traces and refusal traces (resp.) of $P$.

The efficiency preorder in~\cite{CVJ00} is timed variation of the testing preorder in \cite{DH84}.
In~\cite{CVJ00}, (timed) tests are pairs $(O, D)$ where $O$ is a test environment (or user
behaviour, i.e.\ a process that contains $\omega$) and $D \in \nat_0$ is an upper time bound. A
process $P$ satisfies a timed test $(O,D)$ if each discrete trace $v \in \DL(P
\,\|_{\mathbb{A}\backslash \omega}\, O)$ whose duration (i.e.\ its number of 1's) is greater than
$D$ contains some $\omega$. We say that $P$ is faster than $Q$ (written $P\sqsupseteq Q$) if $P$
$d$-satisfies {\it all tests} that $Q$ $d$-satisfies. Moreover, $\sqsupseteq$ can be characterised by
inclusion of refusal traces. This efficiency preorder is qualitative in the sense that a test is
either satisfied or not, and that a process is more efficient than another or not. However, as shown
in~\cite{CV05}, it can be rephrased in terms of a (quantitative) {\em performance function} $p(P,O)$
that gives the worst-case time that $P$ needs to satisfy the test $O$. In more details, 
$P\sqsupseteq Q$ iff $p(P,O) \leq p(Q,O)$ for all test process $O$. Yet, the performance function
(as the preorder $\sqsupseteq$) contrasts processes w.r.t. any possible test environments. In some
cases this might be too demanding and one can make some reasonable assumption about the user
behaviours. In particular, one could be interested in users that have a number of requests (made via
an $in$-action) that they want to be answered (via an $out$-action) as fast as possible. This is the
class of users ${\cal U} = \{U_n\,|\, n \geq 1\}$ where $U_1 \equiv
\underline{in}.\underline{out}.\underline{\omega}$ and $U_{n} = U_{n-1} \; \|_{\{\omega\}}\;
\underline{in}.\underline{out}.\underline{\omega}$ (for any $n>1$). Given these users, the {\em
response performance} is defined to be the function $rp: \nat \rightarrow \nat_0$ such that $rp_P(n)
= p(P, U_n)$ ( $n$ is the number of requests of the user). 

Below we briefly describe how this response performance function is calculated in~\cite{CV05}. To
this aim we only consider the so-called {\em response processes}, i.e.\  processes that can
reasonably serves users in ${\cal U}$\footnote{In~\cite{CV05} a response process is a process that
only perform $in$'s and  $out$'s as visible actions and never produce more responses than
requests.}. Now, we first observe that, for any given $n$, $rp_P(n)$ is obtained as the supremum of
durations of all discrete traces in $\DL(P\;\|\; U_u)$ that do not contain $\omega$. Traces in
$\DL(P\;\|\; U_u)$ are just paths in $\RTS(P \;\|\; U_u)$ that only contain full time steps.
Moreover, for each of such paths there is a corresponding path in $\RRTS(P)$\footnote{This is a
reduced version of $\RTS(P)$. See~\cite{CV05} for more details.} with the same number of conditional
time steps. Thus, to calculate $rp_P(n)$ it will suffice to consider path in $\RRTS(P)$. A first
result in~\cite{CV05} states that the response performance of a response process $P$ is the supremum
of the number of time steps taken over all paths in $\RRTS(P)$ with enough $in$'s and $out$'s to
satisfy the user $U_n$ (so called {\it n-critical paths}). A this stage, a key observation
is that, when the number $n$ of requests is large compared to the number of processes in $\RRTS(P)$,
a $n$-critical path with many time steps must contain cycles. Thus, it turns out to be essential to
find the worst cycles. In~\cite{CV05} these worst-cycles are distinguished to be either {\it
catastrophic} or {\it bad} cycles. A cycle in $\RRTS(P)$ is said to be catastrophic if it has a
positive number of time steps but no $in$'s and no $out$'s. More intuitively, if $\RRTS(P)$
contains a catastrophic cycle, there is at least a path in
$\RRTS(P)$ with arbitrarily many time steps and, hence, there is at least an $n$ such that $rp_P(n)
= \infty$. If $P$ is free from such cycles, $rp_P(n) = an + \Theta(1)$ is asymptotically linear (see Theorem 16
in~\cite{CV05}). The {\it asymptotic factor} $a$ of $rp_P(n)$ is determined
by considering cycles reached from $P$ by a path where all time steps are full and which themselves
contain only time steps that are full; let the {\em average performance} of such a cycle be the
number of its full time steps divided by the number of its $in$'s. We call a cycle {\em bad} if it
is a cycle of maximal average performance in $\RRTS(P)$. Finally, the asymptotic factor of $P$ is
the average performance of a bad cycle. 
 
\section{Performance evaluation with $\fase$}
\fase\ is a useful tool developed at University of Camerino to automatically evaluate the worst-case
efficiency of asynchronous systems. It is written in Java and consists of two main components. The
former one is the parser unit that reads a string representing a PAFAS process $P$ and builds its
$\RTS(P)$. The second component is the performance unit the uses the $\RTS(P)$ to implements all the
technical stuffs discussed in the previous section. Moreover, it also provides some
diagnostic informations that help the user to better understand to behaviour of the process.

The tool automatically checks if a process has some catastrophic cycles or not. The
original solution proposed in \cite{CV05} makes use an algorithm whose a complexity is
$\theta(N^3)$ ($N$ are the nodes of the graph $\RRTS(P)$ of a process $P$). If $P$ is a complex
process, the state space of $\RRTS(P)$ can be very large and the original solution becomes slow.
~\fase\ (see~\cite{BCC+09}) adopts a new solution that takes advantage from the correspondence
between cycles and strongly connected components \cite{Aho83}. This improved solution has a complexity of
$O(N + E)$ where $N$ and $E$ are the nodes and the edges in $\RRTS(P)$. We refer to \cite{BCC+09}
for a running time comparison between the two algorithms.

If $P$ does not have catastrophic cycles, $\fase$ looks for bad cycles in order to determine its
average performance. In doing that, $\fase$ adopts the original solution \cite{CV05} with some
improvements that provide the user with information about the bad cycle just computed. 
Since bad cycles are computed in $O(N^3)$, we are currently investigating new strategies to limit in
some way this complexity.

We are also working on a solution to determine the response performance of $P$ for a 
given $n$. Different approaches are under investigation but they still need to be validated. 
Currently, $\fase$ executes an exhaustive search on $\RRTS(P)$ that looks for the $n$-critical 
path whose duration is maximal; clearly as $n$ increases this solution becomes soon intractable, 
especially for complex processes. 

\section{A Case Study and concluding remarks}
In \cite{BCC+09}, \fase\ has been used to evaluate the worst-case efficiency of three different
implementations of a bounded buffer of capacity $N+2$ whit $N \in \mathbb{N}^+$. These
implementations have already been considered in~\cite{CDV01}. We were interested in studying if the
results steted in~\cite{CDV01} still hold in our qualitative setting. \fifo\ is a bounded-length
first-in-first-out queue, purely sequential and without overhead (in terms of internal actions).
\pipe\ implements the buffer as the concatenation of $N+2$ cells, where each one is an I/O device
that stores at most one value. Cells are connected end-to-end that is the output of a cell is the
input of the next one. Finally, \buff\ uses $N$ cells as a storage \mem\ that interacts with a
centralised buffer controller $\bc$; \bc\ manages \mem\ in a circular fashion and also retains the
oldest undelivered value and outputs it whenever possible. In \cite{BCC+09} we have obtained
interesting results relating the three buffers. We have used \fase\ to prove that none of these
implementations has catastrophic cycles. Moreover, we have also shown that $rp_{\fifo}(n) = 2n$,
$rp_{\pipe} = 2n + N + 1$ and $rp_{\buff}(n) = 4 n$. Thus, \fifo\ is more efficient than both
\pipe\ and $\buff$, while \buff\ is more efficient than \pipe\ iff $n \leq \lfloor N+1/2 \rfloor$.
These results are quite different from those in~\cite{CDV01} where the buffers have been
compared by means of the efficiency preorder in~\cite{CVJ00}. The authors proved that \fifo\ and
\pipe\ (but also \buff\ and $\pipe$) are unrelated (i.e. the former process is not more efficient
than the latter and vice versa) while \fifo\ is more efficient than \buff\ but not vice versa.
Intuitively, this is due to the fact that $rp$ contrasts processes w.r.t. to a specific class of
user behaviours while the preorder $\sqsupseteq$ contrasts process w.r.t. any possible test. To
prove if our intuition is correct, we are working on the definition (and characterisation) of a
slight variation of the faster-than preorder given in~\cite{CVJ00} that allows us to contrast
processes only w.r.t. user behaviours by some variant of refusal trace inclusion. Moreover, it still 
remains to investigate in 
which extent the approach described in~\cite{CV05} to other possible scenarios and to a different
(maybe larger) class of tests. For what concerns $\fase$, a first important result
achieved in \cite{BCC+09} is the improvement of the catastrophic cycles detection since ensuring
their absence is the basis for any further performance analysis. For bad cycles, we
are obtaining encouraging results but they  are still under validation. Moreover, it's still open
the problem of finding the $n$-critical path for a given $n$; we believe that further studies on
the characteristics of an $n$-critical path can help us to find a useful solution.

%
%
%
%
\end{document}